\documentclass[10pt,letterpaper]{article}
\usepackage[top=0.85in,left=2.75in,footskip=0.75in]{geometry}

\usepackage{amsmath,amssymb}

\usepackage{changepage}

\usepackage[utf8x]{inputenc}

\usepackage{textcomp,marvosym}

\usepackage{cite}

\usepackage{booktabs} 
\usepackage{subfigure}
\usepackage{adjustbox}
\usepackage[ruled]{algorithm2e} 
\usepackage{tabularx,ragged2e,booktabs,caption}
\usepackage{nameref}
\usepackage{varioref}
\usepackage{cleveref}
\RequirePackage[textsize=scriptsize]{todonotes}

\usepackage{microtype}
\DisableLigatures[f]{encoding = *, family = * }

\usepackage{array}

\newcolumntype{+}{!{\vrule width 2pt}}

\newlength\savedwidth

\raggedright
\usepackage[aboveskip=1pt,labelfont=bf,labelsep=period,justification=raggedright,singlelinecheck=off]{caption}

\makeatletter
\renewcommand{\@biblabel}[1]{\quad#1.}
\makeatother

\date{}

\usepackage{lastpage,fancyhdr,graphicx}
\usepackage{epstopdf}
\fancyheadoffset[L]{2.25in}
\fancyfootoffset[L]{2.25in}
\begin{document}
\vspace*{0.2in}

\begin{flushleft}
{\Large
\textbf\newline{On Good and Bad Intentions behind Anomalous Citation Patterns among Journals in Computer Sciences} 
}
\newline
\\
Joyita Chakraborty\textsuperscript{1},
Dinesh Pradhan\textsuperscript{1},
Hridoy Sankar Dutta\textsuperscript{2},
Subrata Nandi\textsuperscript{1},
Tanmoy Chakraborty\textsuperscript{2*}
\\
\bigskip
\textbf{1} Dept. of Computer Science \& Engg., NIT Durgapur, West Bengal, India
\\
\textbf{2} Dept. of Computer Science \& Engg., IIIT Delhi, New Delhi, India
\\
\bigskip

%
%





* tanmoy@iiitd.ac.in

\end{flushleft}
\section*{Abstract}
Scientific journals are an important choice of publication venue for most authors. Publishing in prestigious journal plays a decisive role for authors in hiring and promotions. It also determines ranking and funding decisions for research groups, institutions and even nations. In last decade, citation pressure has become intact for all scientific entities more than ever before. Unethical publication practices has started to manipulate widely used performance metric such as ``impact factor" for journals and citation based indices for authors.  This threatens the integrity of scientific quality and takes away deserved credit of legitimate authors and their authentic publications. 

In this paper we extract all possible anomalous citation patterns between journals from a Computer Science bibliographic dataset which contains more than 2,500 journals. Apart from excessive {\em self-citations}, we mostly focus on finding several patterns between two or more journals such as {\em bi-directional mutual citations}, {\em chains}, {\em triangles}, {\em mesh}, {\em cartel relationships.} On a macroscopic scale, the motivation is to understand the nature of these patterns through weighted directed graph which models how journals mutually interact through citations. On microscopic level, we differentiate between possible intentions (good or bad) behind such patterns. We see whether such patterns prevail for long period or during any specific time duration. For abnormal citation behavior, we study the nature of sudden inflation in impact factor of journals on a time basis which may occur due to addition of {\em irrelevant} and {\em superfluous} citations in such closed pattern interaction. We also study  possible influences such as abrupt increase in paper count due to the presence of self-referential papers or duplicate manuscripts, author self-citation, author co-authorship network, author-editor network, publication houses etc. The entire study is done to question the reliability of existing bibliometrics, and hence, it is an urgent need to curtail their usage or redefine them.




\section*{Introduction}
\label{S:1}

Are existing quality metrics reliable any more to measure the importance of scientific entities (journals, conferences, researchers, institutions etc.)? Colossal expansion of bibliographic resources since 2000 has now directed prime focus on quality assessment. This is due to several reasons such as increased author and publication rate, digital accessibility, visibility, peer-peer collaboration propensity, open choices of publication venue etc.  
Several factors such as rapid increase in number of publications per journal along with multiple choices of publication venue, open digital access, lowering subscription charges, increased visibility and collaboration has scattered inter-communication networks amongst various scientific entities (authors, papers, journals, institutions etc.). However, this has radically changed in recent years with introduction of impact factor as quality metric for journal \cite{Seglen:2003, Garfield:1999,PradhanPMNC17} along with upsurge of other author performance indices like h-index, g-index etc. for strict assessment of research quality. More than ever before tremendous publication and citation pressure prevails in the academic community. More publications in high impact factor journal could bring unprecedented reward for authors. For editors and publishers, it could increase popularity and attract more citations for their journals. Henceforth, in recent times following unethical practices journals and their authors, editors, publishers have started to artificially boost impact by manipulating citation and paper count. Benefits are astonishingly high and the effort required to do so is almost nothing. 

Easiest way is by adding excessive `self-citation'. Since 2009, Thomson Reuters (Currently known as Clarivate Analytics) in annual journal citation report consisting of journals they index, started reporting and blacklisting those journals which were excessively self-cited and involved in other anomalous citation patterns such as ``Citation Stacking" \cite{ThomasReuters,analytics2017title}. They have also started to calculate impact factor for journals they index by removing self-citation. {\em Impact Factor} (IF) for journal refers to ratio between total incoming citation received by that journal for preceding two (or five) years to total paper count considering only those articles which are published in that same time window. Majority of existing research in literature focus on biasness in self-citation \cite{DagAksnes:2003,DagAksnes:2007,Glanzel:2004} while in reality, along with self-citation there exists some other anomalous patterns where a superfluous citation exchange occurs between two or more journals such as ``Citation Stacking" and ``Citation Cartels" \cite{Fister:2016}. Cases of journal misconduct are reported where authors, editors and publishers for their mutual benefit shake hands to raise overall impact \cite{arnold2009integrity}. This leads to growing concern towards downfall of research quality. Moreover, an urgent need is to identify and eliminate beforehand such grouping of anomalous citation patterns. Redefining strict bibliometrics is necessary.  

Large number of citations between two journals does not always imply a bad intention. But, there is a need to draw line between acceptable and unacceptable citation behavior. Our motivation is intrigued from a basic question -- {\bf \em In generalized terms by looking at a citation pattern, could we identify it as anomalous, or whether we could do it by adding some characteristic feature? What are some fundamental reasons behind such patterns and could it help us differentiate between good and bad intentions? Are such patterns domain and time specific?} Earlier also such citation anomalies existed but due to lack of resources (datasets and computational advances) they could not be spotted. However today with easier options for archiving large dataset and evolution of bibliometrics as a new field, scope for large-scale analysis and detection of such patterns has become relatively easier. 

Here, we conduct our experiment on Computer Science bibliographic dataset collected from Microsoft Academic Search consisting of more than 2,500 journals which are grouped initially with 490,249 published papers \cite{ChakrabortySTGM13,0002SGM14}. We observe that since 1990, there is 60\% inflation in publication rate along with 74\% increase in citation rate. As a result we narrow down our study to a period of 1990-2012. We mainly extract patterns with bi-directional citation edge between a group of journals and excessive one way citation edge from a formalized weighted directed graph. In particular, we categorize possible anomalous pattern such as {\em Mutual citation}, {\em Citation Chain}, {\em Citation Triangle} and {\em Citation Mesh} along with {\em self-loops} (excessive self citations) and {\em Citation Cartel} (excessive one way citations). On macroscopic level, we analyse these patterns over its entire publication period to model how journals mutually interact. The motive of profit for journals is raising their impact factor. On microscopic level, we study sudden growth in impact factor curve on temporal scale. This helps to identify whether such patterns prevail for long period or during any specific time duration. We attempt to justify between genuine or malicious intent behind such anomalies. For instance, new journals exposing research work in a new or specialised field attracts more self citations. Also, review journal (publishing survey papers) or journal consisting of multiple papers co-authored by many authors is prone to self-citation and mutual citation patterns. Further, interaction of microscopic entities such as influence of publication houses, author co-author relations, author editorial board relations also play a vital role in such macroscopic journal-journal citation anomalies. Our main objective is to define a spectrum and its characteristics so as to label a pattern detected as an anomaly.

\subsection*{Major findings} (i) Over entire publication age, if a pair or group of journal do not exchange large citation to each other then on temporal impact factor analysis we find such journals do not mutually support each other to inflate impact factor.  (ii) In many pair of journals due to mutual citations we find sudden peak in impact factor and then a constant rise . In such cases, microscopic study of measuring increase in paper count and author citation study during that specific time period is necessary. (iii) Abrupt increase in paper count could be an alarming indication. (iv) Also, characteristic feature of sudden peak in impact factor is domain and time specific. (v) It is also influenced by nature of interaction between publication houses and thus leads to {\em citation triangle} and {\em citation mesh} patterns. IEEE and Elsevier publishers contribute highest towards such pattern. (vi) Often, we find that review or sister journal from either same or different publisher mutually inflate impact factor of its parent (older) journal in same domain. Grouping between more than two journal in citation chain pattern is characteristic of such behavior. (vii) Newly published or less visible journals are prone towards {\em self-citation} and {\em citation cartels}. (viii) Author self-citation by adding self-referential papers is one of fundamental reasons behind such pattern. (ix) We find author editorial board relations which influence large mutual citations. (x) High weighted mutual citation cases are significant for our study whereas, low and medium weighted mutual citation cases are less prone towards anomaly. Even when we see sudden peak in impact factor in such cases, they mostly occur for entire publication age. They are neither time specific nor paper count increases abruptly for them.

\section*{Background}
\label{S:2}

Impact factor is widely used bibliometric for journals. Unfortunately, due to its distortion by unethical publication standards, sham peer review, addition of erroneous, self-referential and duplicate manuscripts, excessive self-citations and systematically formed mutual citation grouping between author, editor and publishers; it could neither be directly correlated with quality of publications in that journal nor account credibility for authors of those papers \cite{gasparyan2016statement,wilhite2012coercive,krell2010should, HicksD:2015}. Besides publishers and editors, authors are also prime beneficiaries of a high impact factor journal because it biases and collects large citations for them anyway. In an article by Douglas N. Arnold, several cases of both author and journal misconduct have been reported \cite{arnold2009integrity}. 

Earlier, many works have studied kinetics of journal self-citations and how they can be manipulated to artificially inflate impact factor. Chorus and Waltman \cite{Chorus:2016}  proposed a measure, called IFBSCP (Impact Factor Biased Self-Citation Practices), a ratio between share of self-citation count to papers published particularly in years taken into consideration the impact factor calculation (i.e., last two years) to relative share of self-citations to papers published in that journal for preceding five years. Stephen M. Lawani \cite{Lawani:1982} discussed about different classes of self-citations, namely diachronous and synchronous self-citations.  Bai \cite{Bai:2016} discovered the anomalous citations between articles using their collaborative time-factors. Fong \cite{Fong:2017} discussed the problem of Coercive Citation (adding irrelevant citations) and Padded Citation (add unnecessary citations to manuscripts prior to submission).

Cui et al. \cite{CuiJ:2010} showed relationship between researchers by combining multiple networks (paper citation, author citation and author collaboration network) with an aim to improve link prediction in citation network. One newly identified pattern is `citation cartel' where  a group of journals give excessive one-way citation to one particular journal as compared to other journals in same domain. In 1999, this phenomenon first came into light in an essay published by G. Franck \cite{FranckG:1999} who defined it as a class of journals and editors coming together in close association for a short period to mutually inflate impact factor and influence each other's reputation in the scientific community. Mongeon et al. \cite{Blog1} attempted to address issue of `Citation Stacking' and `Citation Cartels'. Fister et al. \cite{Fister:2016} approached towards identifying cartels by finding interlinked relationships in multilayer networks, i.e., paper-paper and author-author citation networks using triplets ``subject-predicate-object" logic in RDF (Resource Description Framework) format. Heneberg \cite{HeneberP:2016} identified a case on citation stacking where three journals of Physics published by same publisher ``Editura Academiei Romane" are continuously giving high number of citations to each other. Also, it is observed that this citation network becomes dense during post publication year 1-2 and but is invisible during the year of publication. Several attempts were made to redefine existing bibliometrics \cite{CareyRM:2016}.

Major difference of our approach as compared to other methods in literature is that only few cases of such anomalies have been reported to exist; but here, we study a significantly large bibliographic dataset of more than 2500 journals. Along with self-loops and excessive one way cites, in this paper we come up with various extensions to existing patterns such as \textit{Citation Chain}, \textit{Citation Triangle} and \textit{Citation Mesh}. We also study whether such patterns are only visible on a temporal scale or global perspective. Further, we find reasons behind these anomalies, i.e., impact of publication houses, domain specific nature, influence of scholarly relations such author co-author relation, author editor relations etc. At last, we  differentiate between good or bad intentions behind such grouping of journals. This to best of our knowledge, is first attempt to detect different anomalous patterns from such a large scale bibliographic dataset.

\section*{Materials and Methods}

\subsection*{Massive dataset}
\label{S:3}

\begin{figure}
  \includegraphics[width=\columnwidth]{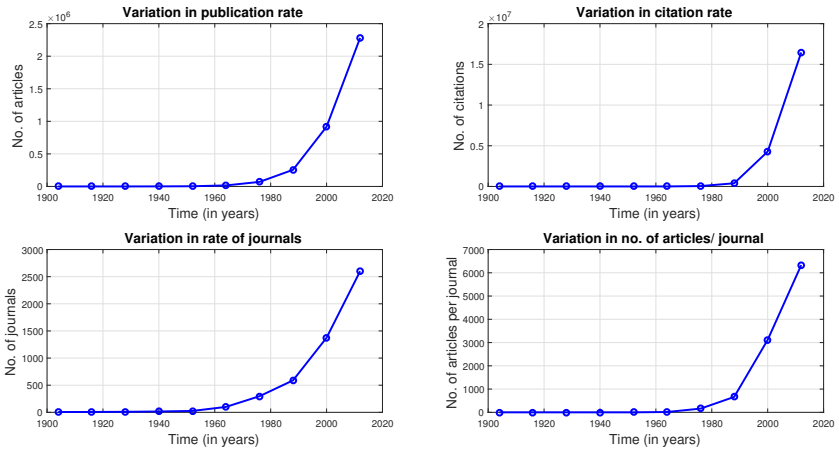}
  \caption{{\bf Growth in the rate of publication, citation, number of journals and number of articles published per journal on a time basis from 1900 till 2012. We observe that since a period of 1990 to 2000, there is rapid inflation in each of these figures. While rate of publication has seen 60\% inflation; citation rate has seen 74\% inflation. Along with the increase in journals, number of articles published per journal has increased two-fold from $3,116$ in 2000 to $6,332$ in 2012.}}
  \label{fig:datasetall}
\end{figure}

Earlier works in literature reported only few cases of such anomalous citation patterns. For instance, Heneberg \cite{HeneberP:2016} detected excessive self-citation and citation stacking by analyzing only 60 journals. Here, 
we crawled one of the largest available bibliographic datasets in Computer Science domain, ``Microsoft Academic Search" (MAS) which includes as of 2012, fetched 2,101,548 papers with unique paper index and corresponding bibliographic attributes such as publication year, venue, author, citation list etc. We find total number of publications with at least one citation as $1,088,452$ whereas, number of authors and citation include $2,662,300$ and $16,463,489$ respectively \cite{Chakraborty18,DeyRCG17}. We restrict our study only to journals and a publication period ranging between 1990-2012. Total number of journal papers include $490,249$ ($45.04\%$) and corresponding journal count is $2,621$. In 2000, publication rate has seen $60\%$ inflation along with $74\%$ increase in citation rate (see Figure \ref{fig:datasetall}). Number of articles published per journal has increased two-fold from $3,116$ in 2000 to $6,332$ in 2012. Number of authors per paper is $2.48$ whereas, number of papers per author is $5.17$. 

\subsection*{Data pruning and filtering into resultant citation graph}

Our prime objective is to study intention behind influential citation edges and detect suspicious citation anomalies from a large journal to journal directed citation graph. As a result in the beginning we aim to filter out all such edges which do not make significant contribution to our graph. For example, journals receiving one citation in entire publication period. 

Initially we form a paper-paper directed citation graph 
with papers receiving at least one citation. Next, we group those papers based on respective journals they are published in. The resultant journal-journal directed citation graph consists of 2,621 vertices. To illustrate, when papers P\textsubscript{1}, P\textsubscript{2},...P\textsubscript{n} published in journal $J_{i}$ cite papers Q\textsubscript{1}, Q\textsubscript{2},...Q\textsubscript{m} published in journal $J_{k}$, a directed edge is drawn from $J_{i}$ to $J_{k}$. Weight of edge between them is total number of outgoing citations ($x_{ik}$) from all papers of $J_{i}$ pointed to $J_{k}$. From this graph, it is evident that for a single vertex, 3 directed edges exist -- \textit{self-directed loop}, \textit{incoming edge} and \textit{outgoing edge.} In order to understand characteristic properties of anomalous citation behavior, it is essential that we take into account contribution made by a vertex in all directions equally. 

\begin{figure}
\centering
  \includegraphics[scale=0.4]{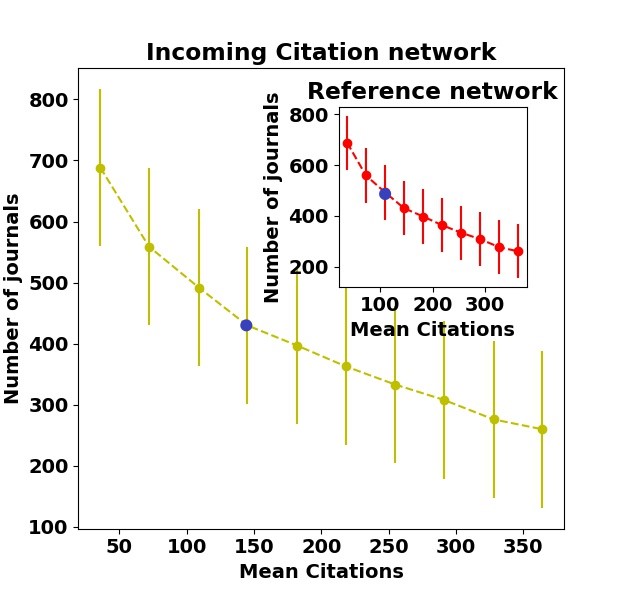}
  \caption{{\bf Variation in number of journals with the mean citations calculated for the entire journal to journal citation graph. We choose those journals into our study which obtain greater than 110 citations, (i.e., 60\% of mean in incoming citation graph); and greater than 109 citations (i.e., 40\% of mean in outgoing reference graph). Remaining number of journals in the resultant graph are $428$ and $559$ respectively. We find that only journals which collect greater than 100 citation in its entire publication period could be further, studied  on a time basis for citations collected by that journal in impact factor time window,  and hence we choose such a threshold.}}
  \label{fig:meancitationj}
\end{figure}

Next, non-contributing edges are filtered out. Individually for each journal, we calculate mean citation collected or referred by a journal over its effective citation age (considering only those years when a journal has received or given at least one citation). Only those citation edges remain in graph which has weight greater than mean. A shortcoming observed is when a journal receives lower mean citation over its publication period then all of its corresponding edges in graph become insignificant for our study. Hence, we recalculate mean citation for entire journal-journal graph. Mean citation for incoming citation graph and outgoing reference graph is obtained as 277 and 182 respectively. Standard deviation ($\sigma$) for vertex count in incoming citation graph is $106.97$ and outgoing reference graph is $128.75$. For related reasons, we relax mean by $\sigma$ thereby taking journals with citation count 60\% of mean ($>$ 110 citations) in incoming citation graph and 40\% of mean ($>$ 109 citations) in outgoing reference graph (see Figure \ref{fig:meancitationj}). Moreover, citation distribution on time basis for impact factor year is feasible to study only if weighted citation edge has value greater than 100. We notice a drastic fall in vertex count with 428 and 559 from initial 2,621 vertices respectively in resultant incoming citation and outgoing reference graph.

All calculation for time basis study of paper and citation count is done considering impact factor time duration. For instance, paper count for a journal in year $y$ is calculated as recently published articles in that journal during preceding two years ($y-1$) and ($y-2$). Citation count is total citation collected by recently published articles of that journal during same preceding two years.

\begin{table*}[htb]
\centering
\tiny
\resizebox{\textwidth}{!}{
\begin{tabular}{l l}
\hline
\textbf{Journal Title} & \textbf{Notation} \\
\hline
Journal of The Acoustical Society of America & JASA \\
ACM Sigcse Bulletin & ASB\\
Environmental Modelling and Software & ENVSOFT\\
IEEE Transactions on Energy Conversion & ENCONV \\
IEEE Transactions on Information Theory & TIT \\
Computing Research Repository & CoRR \\
ACM Special Interest Group on Programming Languages & SIGPLAN \\
Journal of the American Society for Information Science & JASIS \\
ACM Transactions on Programming Languages and Systems & TOPLAS \\
IEEE Transactions on Communications Home & TCOM \\
IEEE Journal on Selected Areas in Communications & JSAC \\
IEEE Communications Magazine & ComMag \\
IEEE Communications Letters & ComLet \\
IEEE Transactions on Pattern Analysis and Machine Intelligence & PAMI \\
Bioinformatics/Computer Applications in The Biosciences & BIOINFORMATICS \\
International Journal of Computer Vision & IJCV \\
Pattern Recognition & PR\\
Pattern Recognition Letters & PRL \\
ACM Sigarch Computer Architecture News & SIGARCH \\
IEEE Transactions on Image Processing & TIP \\
Applied Mathematics and Computation & AMC \\
BMC Bioinformatics & BMC \\
Computer Vision and Image Understanding & CVIU \\
European Transactions on Telecommunications & ETT \\
IEEE Transactions on Signal Processing & TSP \\
IEEE Transactions on Wireless Communications & TWC \\
Scientometrics & Sciento \\
Computer Networks & CompNetw \\
Computer Communication Review &  CCR\\
Computer Communications & ComCom \\
Journal of The American Society for Information Science and Technology & JASIS \\\hline
\end{tabular}}
\caption{{\bf Description of journals and their notation used}}\label{tab:b}
\end{table*}


\subsection*{Temporal impact factor analysis}
\label{S:4}

\begin{figure}[!t]
\centering
  \includegraphics[scale=0.4]{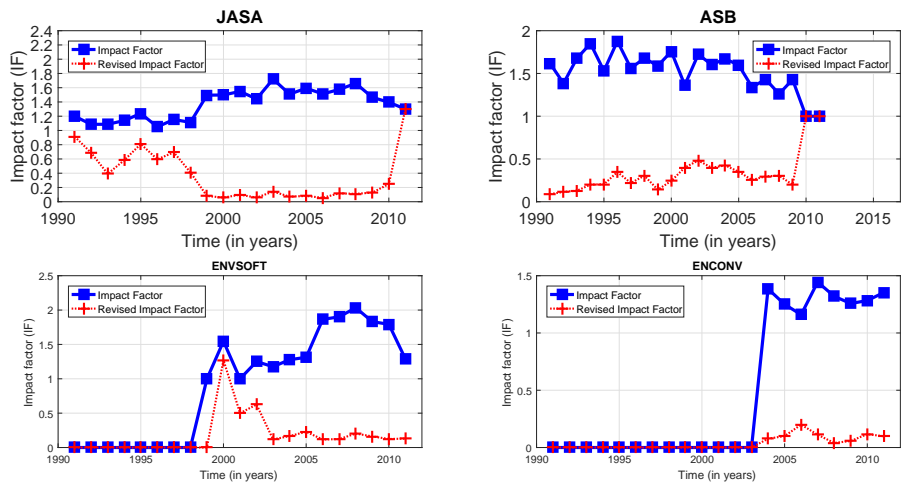}
   \caption{{\bf Variation in impact factor and revised impact factor on a time basis of top four excessively self-cited journals --  \texttt{JASA}, \texttt{ASB}, \texttt{ENVSOFT} and \texttt{ENCONV}. Here, revised impact factor is calculated after removing self-citation. We see that revised impact factor curve almost decreases to 0 for all four journals. While \texttt{ENVSOFT} is a new journal which gets excessive self-cites after 5 years of publication, \texttt{JASA} is amongst few journals published in a specialized field of ``Acoustics". It is also highest self-cited journal (43,383) in our dataset. \texttt{ENCONV}, after 2003, gets sudden peak in impact factor whereas revised impact factor curve for it shows a sharp decrease.}}
  \label{fig:exselfcite}
\end{figure}

The idea of profit behind a group of journals coming in close nexus is raising impact factor \cite{0002N16a}. On global perspective by studying citation trajectory of journals over entire publication period, large number of mutual citations between a pair or group of journals could not be readily claimed as anomalous. A common characteristic feature behind such pattern is sudden inflation in impact factor curve.
In recent years, excessive self-citation is considered a bad gateway towards achieving high impact factor. Hence, {\em Revised Impact Factor} ({\em RIF}) is calculated by removing self-citation contribution from total citation count collected by that journal for preceding two years \cite{ThomasReuters}. Pre-analysis of our dataset reveals that since 1990's, self-citation has abruptly increased in huge volume due to expansion of research in Computer Science domain in late 1990\'s and early 2000 \cite{chakraborty2015categorization,ChakrabortyN18}. Earlier to that self-citation was the only way to make journals visible to academic community and it could not be considered as done with bad intention. Few journals in our dataset also receive as high as 80\% to 90\% self-citations. For instance, \texttt{JASA} (see Table \ref{tab:b} for abbreviation of the journals) receives 85.71\% self-citations in its entire publication age (see Figure \ref{fig:exselfcite}). 

Here, we mainly focus on studying impact factor curve on a time basis for bi-directional mutual citation cases. Citation abnormalities in such patterns are difficult to detect. Though impact factor came in use since 1975, its effect after 2000 to 2005  is more prominent due to digitalisation, visibility and wider accessibility of entities. We begin with a basic question -- \textbf{\textit{ Could sudden mutual exchange of citations to recently published articles be an indication of possible anomaly?}} 

\begin{figure}
\centering
   \includegraphics[width=\columnwidth]{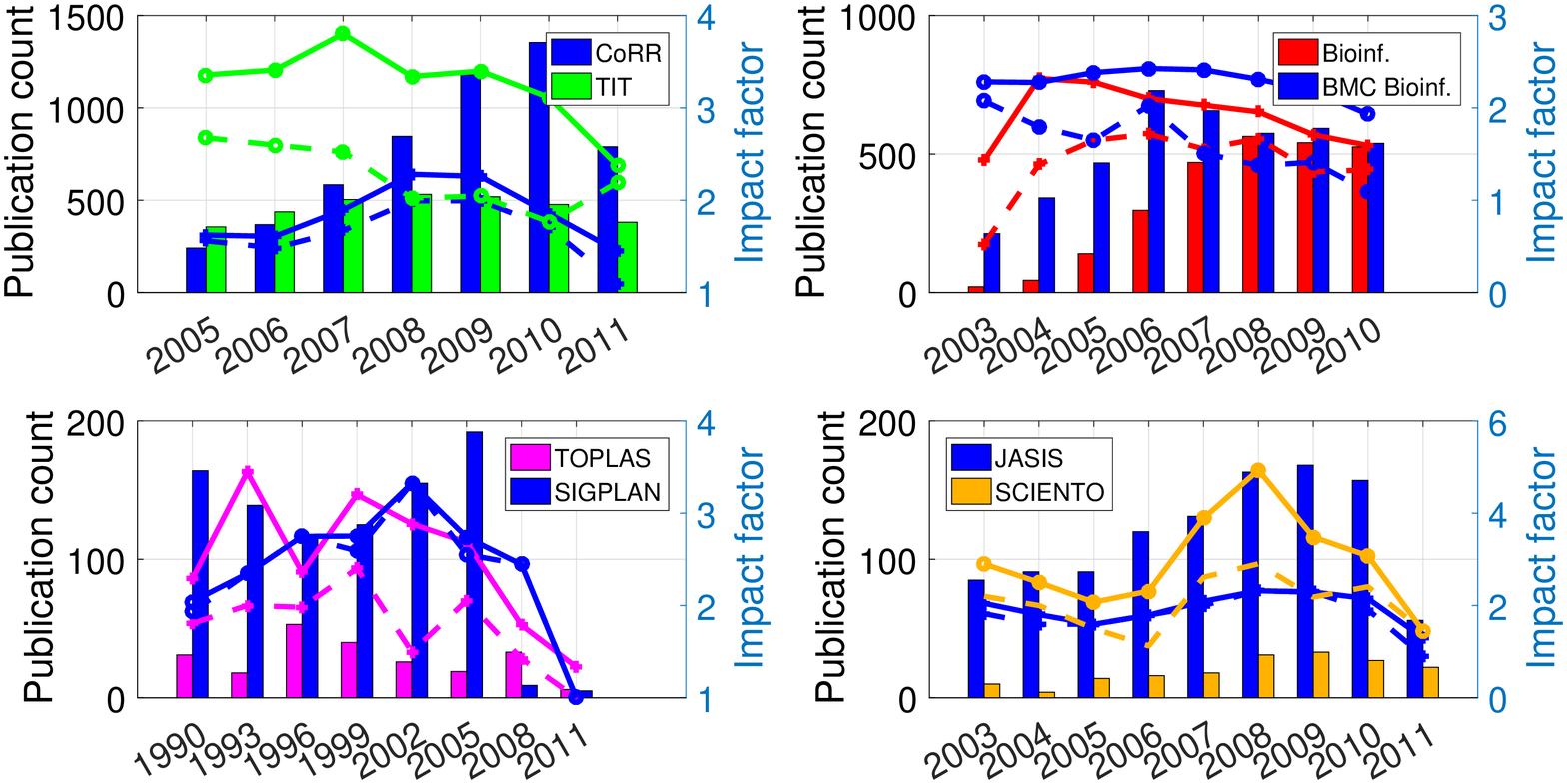}
   \caption{{\bf Temporal impact factor study of four mutually citing pair of journals out of which \texttt{CoRR} and \texttt{TIT}, \texttt{BIOINFORMATICS} and \texttt{BMC Bioinformatics} belong to {\em high weighted mutual citation bucket}, \texttt{TOPLAS} and \texttt{SIGPLAN} belong to {\em medium weighted mutual citation bucket} and \texttt{SCIENTO} and \texttt{JASIS} belong to {\em low weighted mutual citation bucket}. Bar graph refers to publication count of journals on time basis whereas, line graph and dashed line graph depict variation in impact factor and revised impact factor on time basis respectively. Here, revised impact factor is calculated after removing citations collected from mutually cited journal for preceding two years to recently published articles. Since, mutual citation patterns are time specific we only plot those data points where impact factor has largely inflated due to mutual interaction.}}
  \label{fig:mutualcite}
\end{figure}

It is likely that authors will cite older, popular, more visible and relevant journals to their work but we also could not completely rule out the possibility of anomalous citation activities in such journals. For instance, \texttt{TIT} is an old journal first published in 1953. A sudden citation exchange occurred between \texttt{TIT} and \texttt{CoRR} for a specific time period between 2005 to 2011 which is unusual. Although both journals are mutually citing each other however, contribution made by \texttt{CoRR} to \texttt{TIT} is dominant. Thus biasing journal is \texttt{CoRR} and biased journal is \texttt{TIT}. Another alarming indication is that paper count of biasing journal i.e., \texttt{CoRR} increases rapidly during 2007 to 2011 with a maximum of 55.31\% references to \texttt{TIT} in year 2009. In return, \texttt{CoRR} receives 26.94\% citations from \texttt{TIT}. When {\em RIF} for \texttt{TIT} is calculated by removing citations given by \texttt{CoRR} in impact factor window, it shows a sharp declining curve (see Figure \ref{fig:mutualcite}). Also, \texttt{CoRR} is a newer journal than \texttt{TIT}. 
Similar case is observed in another pair of journals belonging in same domain -- computational biology \texttt{BIOINFORMATICS} and \texttt{BMC Bioinformatics}. \texttt{BMC Bioinformatics} is a new journal established in year 2000. After initial 4 years of publication it suddenly starts giving huge volume of citations  to an older journal in the field \texttt{BIOINFORMATICS}. Contrastingly, both journals are increasing their publication count two fold each year in a period between 2005 to 2011. {\em RIF} calculated for \texttt{BIOINFORMATICS} journal shows sharp decrease whereas for newer journal \texttt{BMC Bioinformatics} the curve monotonically decreases and it is less affected (see Figure \ref{fig:mutualcite}).    

A journal with history of high impact factor will always try to maintain its popularity and impact. Unfortunately, now-a-days to do so journals are compromising with standard publication ethics and accepting huge volume of duplicate and erroneous papers with sham peer review process. When we see above two pair of journals in resultant journal to journal citation graph, mutually citing edges between them contribute highest towards each other than to any other journal in their citation history. Definitely both journals are mutually uplifting each other in some proportion but always newer journal's contribution is dominant. 

Next, we take into account such mutual citation cases whose bi-directional weights are not significantly large. We later put them under {\em medium mutual citation bucket} and {\em low mutual citation bucket}. Initially, we obtain 123 cases of mutual citation. Next, we divide them into 3 weighted buckets -- {\em high, medium and low mutual citation bucket} containing equal number of 41 mutual citation cases each (see Section 5). In such cases we mostly see that journals which are not superficially involved in only inflating impact factor their references reflect relevance and quality. We take two journals \texttt{SCIENTO} and \texttt{JASIS} from {\em low mutual citation  bucket}. In a period between 2006 to 2009, \texttt{JASIS} gives large citation inflating impact factor of \texttt{SCIENTO} though paper count is consistent and \texttt{SCIENTO} does not exchange with same behavior. Similar case is observed between another pair of journals with same publication house \texttt{ACM} -- \texttt{TOPLAS} and  \texttt{SIGPLAN Notices} which mutually cite each other throughout entire publication period. Both journals belong to same topic of interest ``Advances in programming languages and systems". While  \texttt{TOPLAS} is a premier journal, \texttt{SIGPLAN Notices} is an informal monthly publication journal giving review on several conference proceedings and \texttt{SIGPLAN} activities. {\em RIF} calculated for \texttt{TOPLAS} shows a sharp declining curve (see Figure \ref{fig:mutualcite}).

\begin{figure}
\centering
   \includegraphics[width=\columnwidth]{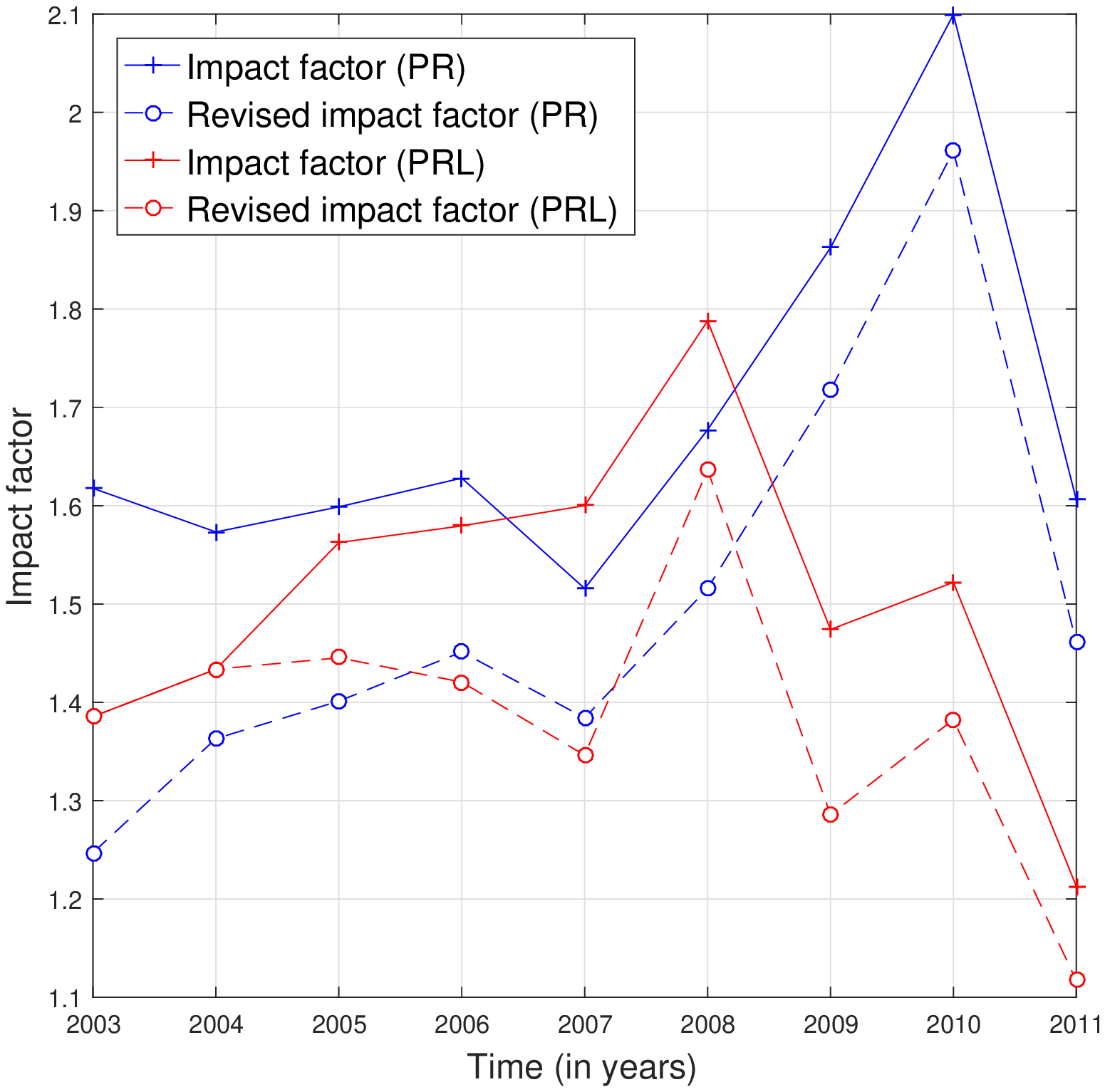}
   \caption{{\bf Temporal impact factor study of mutually citing pair of journal \texttt{PR} and \texttt{PRL}. Line graph and dashed line graph depict variation in impact factor and revised impact factor on time basis for \texttt{PR} and \texttt{PRL} journals respectively. Both sister journals are published by Elsevier and belong to same domain `Pattern Recognition.' Here, revised impact factor is calculated after removing citations collected from mutually cited journal for preceding two years to recently published articles.}}
  \label{fig:mutualcitepr}
\end{figure}

More of such cases are observed in our data set. For instance, two sister journals from same publisher Elsevier `\texttt{PR} and \texttt{PRL}' moderately mutual cite each other. \texttt{PRL} which aims at fast publication of concise review articles on Pattern Recognition has fairly cited its sister journal. For many other cases of mutual citations in medium and low weighted buckets, although there is fair citation exchange over entire publication period its influence on temporal impact factor study completely vanishes with negligible mutual cites (see Figure \ref{fig:mutualcitepr})

Summarizing, we could conclude that sudden mutual exchange of citation between two journals to raise impact factor has become quite an obvious outcome with widespread usage of impact factor. Fault is not in metric itself, rather in ways it could be manipulated by easily adding irrelevant references and publications. Without exhaustive study of underlying reasons behind sudden growth in impact factor; we could not possibly ascertain it as indication towards any suspicious citation activity. Along with it, abrupt increase of paper count in that specific time duration could also be an alarming indication. Another important observation is that such citation behavior is time specific. They could occur at random for any duration. Hence it becomes a challenge to develop automated or generalized algorithms to readily detect such anomalous citation behavior.

\begin{figure}[!h]
  \centering
  \graphicspath{{Figures/}}
  \subfigure[]{ 
  \includegraphics[height=6cm,scale=.28]{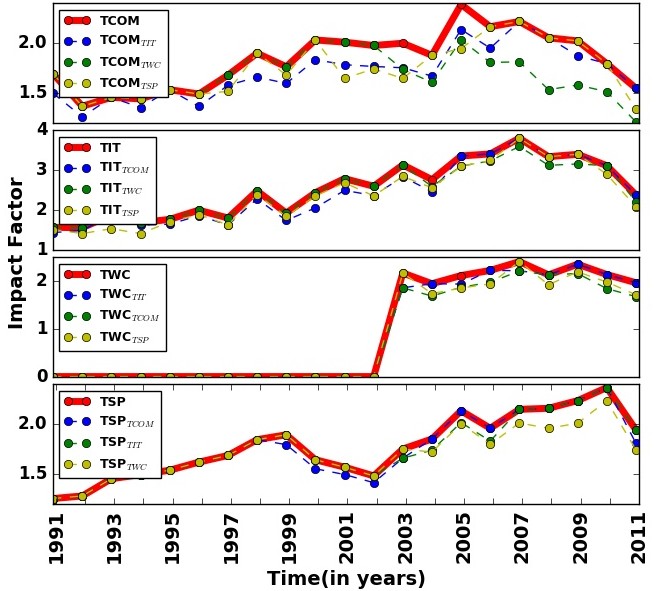}
  \label{fig:mesh_samepubh}
   }
  \subfigure[]{
  \includegraphics[height=6cm, scale=.28]{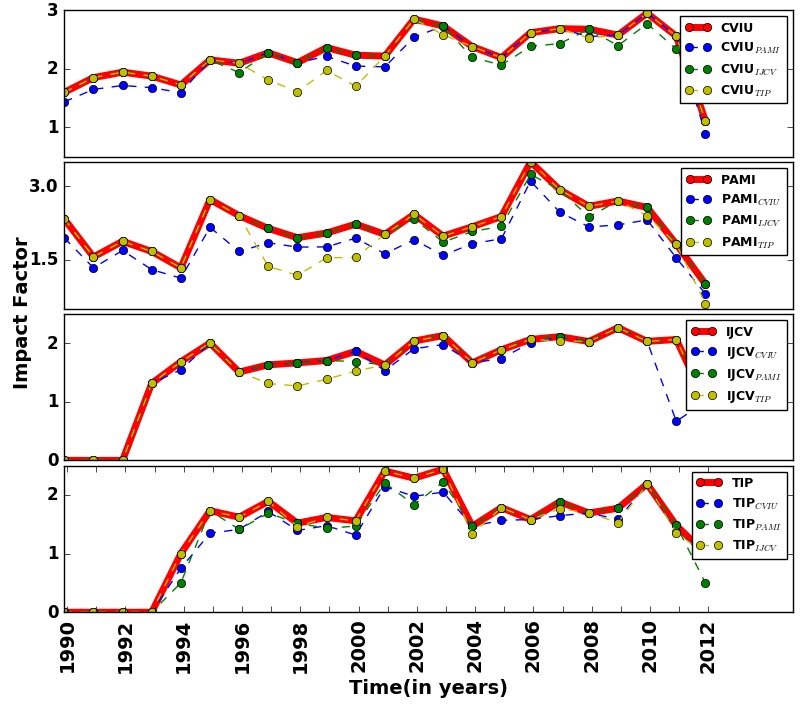}
  \label{fig:mesh_diffpubh}
   }
  \caption{{\bf Study of temporal impact factor of citation mesh pattern. Red line depicts variation in impact factor on time basis, whereas three dotted line refers to revised impact factor after removing citation from other three journals. Figure (a) refers to four journals, namely \texttt{TCOM}, \texttt{TIT}, \texttt{TWC} and \texttt{TSP} belonging to same publication house IEEE. All four journals are mutually citing each other in citation mesh pattern. Figure (b)  refers to four journals, namely \texttt{IJCV}, \texttt{PAMI}, \texttt{CVIU} and \texttt{TIP} belonging to different publication house -- IEEE and Elsevier.}}
\label{fig:citationmesh}
\end{figure}

Bi-directional mutual citation behavior does not necessarily occur only between a pair of journals. We also find such behavior extended to a closed group of three or four journals. For example, all four journals namely \texttt{IJCV}, \texttt{PAMI}, \texttt{CVIU} and \texttt{TIP} belong to same domain and are published by different publication houses. \texttt{TIP} journal published by IEEE and \texttt{CVIU} published by Elsevier which are also reciprocating with large mutual citations make significant contribution towards impact factor of other journals between 1997 to 2000. Notable observations are {\em RIF} calculated for \texttt{IJCV} journal removing citations from \texttt{CVIU} in year 2011 shows a sharp downfall (see Figure \ref{fig:mesh_diffpubh}). Contrastingly, \texttt{PAMI} is one way supported by \texttt{CVIU} for a long period between 1996 to 2010. Interestingly paper count of \texttt{CVIU} and \texttt{TIP} journals rapidly increased during this specific duration whereas, \texttt{CVIU}  published maximum number of articles only during these 4 years of duration. Another closed group of four journals \texttt{TCOM}, \texttt{TIT}, \texttt{TWC} and \texttt{TSP} belonging to same publication house IEEE and same domain ``Communications" mutually cite each other. Out of these four journals, \texttt{TCOM}'s impact factor is inflated to 2.39 by other three journals where, contribution of biasing journals 32.7\% references from \texttt{TWC} and 20.8\% from \texttt{TSP} are maximum.   {\em RIF} calculated for \texttt{TCOM} after removing citations from \texttt{TWC} between 2007 to 2010 depicts a sharp monotonically decreasing curve (see Figure \ref{fig:mesh_samepubh}). Extending it in form of open chain, adding to \texttt{TCOM} we find four journals \texttt{ComLet} $\rightarrow$ \texttt{TCOM} $\rightarrow$\texttt{JSAC} $\rightarrow$ \texttt{ComMag}. All four journals are published by IEEE and belong to same field of research ``Communications". We see that \texttt{ComLet} and \texttt{ComMag} are sister journals featured on a timely manner to publish up to date ongoing research on all aspects of communications including technological and development advances, market trends, upgradation in services and systems, change in regulatory policies and issues whereas \texttt{TCOM} and \texttt{JSAC} mainly focus on telecommunications. Such a nexus could be easily found in journals from same publisher in same field of research where, review journals publish up to date extended content of parent journals which further triggers new research in the field. \texttt{ComLet} is a new journal published in 1997 and there exists a strong citation bond with \texttt{TCOM} since its initial publication year.

Summing up, we observe that three or four journals in group come in close interaction of each other during different time windows to mutually inflate impact factor. This does not mean all bi-directional edges are influential in such groups. In most cases new journals or sister journals from same publisher bias citations of its parent journal. Such groupings are also domain and time specific and largely influenced by publication houses. Temporal impact factor study becomes haphazard as citation exchanges between journals are time specific. We next derive generalized anomalous patterns.

\subsection*{Categorization of possible anomalous citation patterns} 
\label{S:5}
\begin{figure}[!h]
\centering
  \includegraphics[scale=0.4]{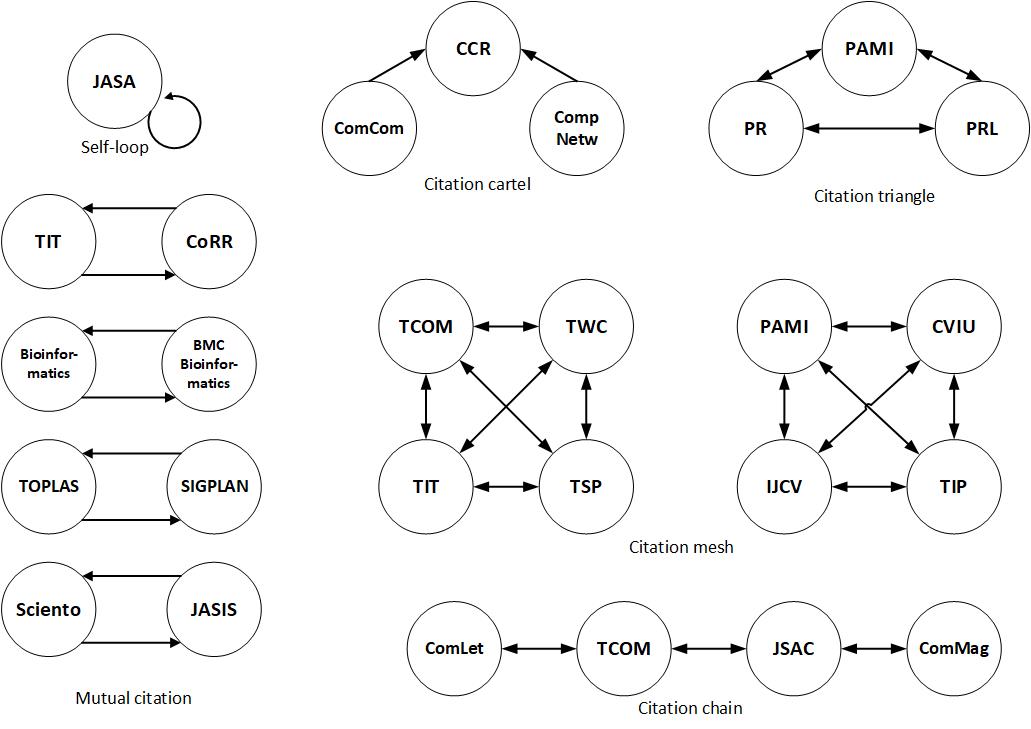}
   \caption{{\bf Different anomalous patterns extracted from journal-to-journal citation graph -- self-loop (excessively self-cited journal), mutual citation patterns and citation cartels (excessive one way citations). We extend mutual citation pattern to citation chain, citation triangle and citation mesh. Circle represents journal and directed edge refers to weighted citation edges. Pleaser refer to Table \ref{tab:b} for the full form of the abbreviations.}}
  \label{fig:patterns}
\end{figure}

Time specific anomalous journal citation activities reflect in common geometrical patterns. From resultant graph, we extract patterns such as self-loop, mutual citations (two vertices with bi-directional weighted directed edge), citation cartels (a group of vertices with uni-directional edges directed towards any single journal). Further, we also extend mutual citations to other shapes such as citation chain, citation triangle and citation mesh. Doing an exhaustive study, we aim to generalize characteristic features exhibited by such patterns. We also see whether it could help us differentiate between good or bad intentions. Through various distinctive properties, by looking at a pattern could we identify it as anomalous? Here we categorize and define possible anomalous citation patterns in journal-journal citation network. Along with it, we discuss methods to extract and analyze such patterns from large dataset (see Figure \ref{fig:patterns}).

\begin{center}

\begin{adjustbox}{max width=\textwidth,center}
\begin{tabular}{ | c | c |}
\hline
\textbf{Name of anomalous patterns} & \textbf{Number of cases} \\
\hline
Self-loops & $361$  \\ 
\hline
Mutual citation &   $123$  \\ 
\hline 
Citation chain &  $56$  \\ 
\hline
Citation triangle & $24$  \\ 
\hline 
Citation mesh &  $11$  \\ 
\hline
Citation cartel &  $43$  \\ 
\hline
\end{tabular}
\end{adjustbox}
\captionof{table}{{\bf Number of cases identified in each of possible anomalous citation patterns.}} \label{tab:patterncasestudy}
\end{center}

\subsubsection*{Self-loop:} It is most commonly found pattern (84.5\%) (see Figure \ref{fig:patterns}). Not all self-loops, but excessive self-citations where a journal gets maximum citation from its own publications and gains high impact factor is anomalous. When weight of self-directed loop becomes greater than 55\% of total citations collected by that journal in its entire publication period, it becomes alarming and excessively self-cited. We choose such a threshold because doing temporal citation study of journals in impact factor time window reveals that mostly after 1995, {\em RIF} for such journals which satisfies this threshold almost decreases to zero. We find 22 journals which are excessively self-cited. Top 5 excessively self-cited journals are \texttt{JASA}, \texttt{ENCONV}, \texttt{ENVSOFT}, \texttt{ASB} and \texttt{CoRR}.

\subsubsection*{Mutual citation:} It refers to a pair of journal with bi-directional weighted directed edges such that they contribute significantly large number of citations to each other rather than other journals in their citation history   (see Figure \ref{fig:patterns}). We obtain a total of 123 cases of mutual citation. It is not always the case that both directed edges are equally influential and hence, to quantify strength of couplets we assign a single coupling weight (w) using Algorithm \ref{algo:couplingweight}. Here, coupling weight (w) refers to a single weight assigned to a pair of journal calculated using Algorithm \ref{algo:couplingweight} from bi-directional mutual citation edges.

\begin{algorithm}
 \SetKwInOut{Input}{input}
 \Input{Weight of bi-directional edges ($x_{ik}$, $y_{ki}$) between two journals $J_i$ and $J_k$ exhibiting mutual citation}
  \uIf{range of ($x_{ik}$, $y_{ki}$) is significantly large ($\sigma >$ mean)}{
    CouplingWeight (w)=max($x_{ik}$, $y_{ki}$)
  }
  \uElseIf{$\sigma$ is sufficiently small}{
    CouplingWeight (w)=mean($x_{ik}$, $y_{ki}$)
  }
\caption{{\bf Computing single coupling weight from bi-directional edges.}}
\label{algo:couplingweight}
\end{algorithm}

We find that if there is huge difference between two weights then larger weight between them becomes coupling weight of two journals. It is because when we calculate impact factor on time basis for the recipient journal, it shows sudden inflation during a specific time duration with large citations contributed by donor journal which is not seen vice-a-versa. On the other hand, if two weights do not exhibit much difference then mean weight calculated becomes coupling weight for that pair of journals. In such cases, we find that both journals on time basis mutually inflate impact factor of each other. 

Based on $CouplingWeight$ (w) as mentioned in Algorithm \ref{algo:couplingweight}, we divide mutual citation cases into three citation buckets with equal number of 41 cases each -- { \em high weighted mutual citation bucket} ($w > 1200$), {\em medium weighted mutual citation bucket} ($w > 450$) and {\em low weighted mutual citation bucket} ($w < 450$) ($w$ is the coupling weight).    We find that characteristic feature of sudden peaks in impact factor is only visible for mutual citations in high weighted mutual citation bucket. Largely they occur due to citation exchange between journals of same publishers and domain specific nature. For majority of cases in medium bucket and low bucket mutual citation, temporal impact factor study reduces down to very less citation exchange between them. For related reasons we thus extend mutual citation cases to a group of journals where all journals mutually cite each other. We aim to study whether in such cases --  medium bucket or low bucket mutual citations make any contribution. We extend mutual citation cases to {\em Citation triangle}, {\em Citation mesh} and {\em Citation chain}.

\subsubsection*{Citation triangle:} Mutual citation occurring between three journals in a closed loop such that if journal $ J1 $ mutually cites $ J2 $ and $ J3 $, then there has to be a bi-directional edge between $ J2 $ and $ J3 $ to form a citation triangle (see Figure \ref{fig:patterns}). We get a total of 24 citation triangles. Strategic considerations between entities such as authors, editors and publishers could be easily hidden behind such patterns. For adjacent vertices in a triangle, we assign a single coupling weight using Algorithm \ref{algo:couplingweight}. On temporal citation study, we find few triangles out of them which form strong nexus and contribute large citations to each other to mutually increase impact factor in which its edges are either in high weighted bucket or in medium weighted bucket. For example, three journals \texttt{PAMI}, \texttt{PR} and \texttt{PRL} from same domain `Pattern Analysis' form a citation triangle. Out of which, \texttt{PAMI} is published by IEEE whereas, \texttt{PR} and \texttt{PRL} are sister journals published by Elsevier. 

\subsubsection*{Citation mesh:} It refers to mutual citation exchange occurring between four or five journals. We can form a citation mesh pattern out of two citation triangles joined by a common edge   (see Figure \ref{fig:patterns}). If journals $ J1 $, $ J2 $, $ J3 $ form a {\em citation triangle} and journals $ J4 $, $ J2 $, $ J3 $ form another {\em citation triangle} and a bi-directional weighted citation edge exists between $ J1 $ and $ J4 $, then it forms a citation mesh pattern. Here, each vertex is joined to every other vertex by bi-directional weighted edge. From mutual citation between two journals to citation mesh pattern, we find few journals which are co-incident and eventually, their grouping increases. Such patterns reveal interesting observations and give us insight on macroscopic scale, how journals mutually interact. For example, four journals \texttt{IJCV}, \texttt{PAMI}, \texttt{CVIU} and \texttt{TIP} belong to different publication houses (IEEE and Elsevier); whereas \texttt{TCOM}, \texttt{TIT}, \texttt{TWC} and \texttt{TSP} belong to same publication house (IEEE). They form citation mesh pattern.  Such patterns are distinctive of how publication house and domain nature influence large mutual citation between a group of journals. We find that this pattern is easily inter-changeable into other patterns on microscopic temporal study.

\subsubsection*{Citation chain:} Citation chain is mutual citation between a group of journals such that all vertices do not have direct bi-directional weighted citation edge between them but, they are connected in form of an open chain (see Figure \ref{fig:patterns}). Although, we obtain 56 chain relations of different lengths 2, 3 and 4 from resultant graph; majority of connecting edges in chain belong to either medium or low mutual citation bucket. Here, length of a chain refers to number of linked edges connecting vertices. If connecting edges of chain are in low mutual citation bucket then such patterns become inappropriate to be studied as anomalous. For example, we find a chain of length 3 between \texttt{ComLet} $\rightarrow$ \texttt{TCOM} $\rightarrow$\texttt{JSAC} $\rightarrow$ \texttt{ComMag}. A notable observation is that in chain if mutually citing edges between adjacent vertices are in high or medium mutual citation bucket then it refers to review or sister journal from either same or different publication houses which is mutually citing either its parent or other sister journals in same domain.

\subsubsection*{Citation cartel:} When a journal receives large number of citation from a group of journals without citing them back at all then such pattern refers to citation cartel -- a group of vertices having uni-directional citation edge towards any single vertex (see Figure \ref{fig:patterns}). We find few examples of such pattern in new and less visible journal which contribute excessive one way citation to an older and more established journal in the field. For example, \texttt{COMCOM} and \texttt{CompNetw} from same publisher (Elsevier) give excessive one way citation to \texttt{CCR} published by \texttt{ACM} all belonging to same domain. \texttt{CompNetw} is also a newer journal.

A generalized observation after extracting such patterns is that if two or more than two journals do not involve in significantly large citation exchange over their entire publication age then such journals could rarely involve in any of the anomalous patterns with a bad intention to mutually inflate impact factor by adding irrelevant references or publications. It is so because prime intention behind a journal getting its hands dirty could only be to gain high impact factor. Throughout our study, we find many such group of journals present in either of generalized patterns extracted above which experience sudden peak in impact factor for a limited time frame largely influenced due to inherent domain nature and publisher's network. A notable characteristic feature of such patterns is that they are domain and time specific. Also, we find that these patterns are interchangeable from global citation study to temporal citation study. For instance, four journals \texttt{IJCV}, \texttt{PAMI}, \texttt{CVIU} and \texttt{TIP}, when we study them globally are mutually involved in citation mesh pattern but when we study time basis impact factor study for a duration of four years \texttt{CVIU} seems to give large citations to the other three journals converting into cartel relationships. 

Now we intend to do an exhaustive study of microscopic entities hidden behind the scenes such as authors, editors and publishers which lets journals do citation gambling in such a sophisticated way.

\subsection*{Micro-level dynamics behind such patterns} 
\label{S:6}
\begin{figure}
\centering
  \includegraphics[scale=0.4]{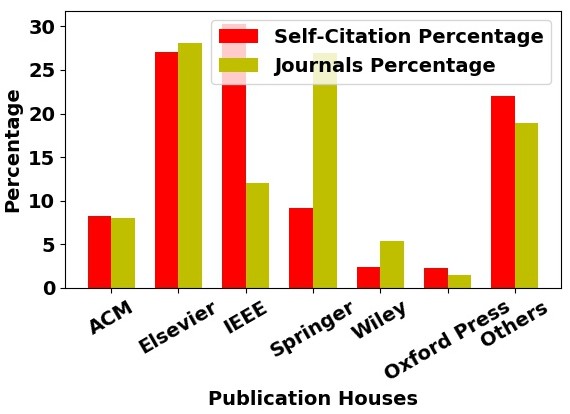}
   \caption{{\bf Six major publication houses such as {\em ACM}, {\em Elsevier}, {\em IEEE}, {\em Springer},{\em Wiley}, {\em Oxford Press} and {\em Others} are represented along x-axis. Green bar graph depict percentage of journals published whereas red bar graph depict percentage of self-citation. Journals published by IEEE and Elsevier are highest self-cited.}}
  \label{fig:selfcitationph}
\end{figure}

Initiating our study to understand how impact factor for journal varies over time, we find many journals in either pair or group which show sudden peak in impact factor and then a constant rise. Revised impact factor calculated for such journal shows a sharp downfall. In derived patterns, we find large mutually citing edges showing more of this characteristic. In addition to that, if biasing journal shows abrupt increase in paper count then it becomes an alarming indication. Microscopic key entities involved are authors, editors and publishers.

A high impact factor journal benefits an author also, because it attracts citations for them anyway. Several instances in literature shows author adding self-referential and duplicate manuscripts and consequently, journal is excessively self-cited. However, there could also be genuine reasons behind excessive self-citation. For instance, \texttt{JASA} is highest self-cited journal (43,383). Reason behind is due to presence of very few journals in this specialized field ``Acoustics." New journal with lesser visibility such as \texttt{ENVSOFT} and journals in specialized disciplines attract excessive self-citations after four to five years of publication. With pressure of gaining high impact factor, it has become a frequently occurring pattern. After Thomson Reuters for journals they index has started to blacklist journals with excessive self-citation and citation stacking patterns; journals are getting more cautious to do such gambling in group.

Author self-citation is also a prime reason behind excessive self-citation for journals. Similarly, for time specific mutual citation pattern we conduct author citation study on time basis to see which authors  contribute maximum towards giving large citation in impact factor time window. A notable observation is that we obtain a set of overlapping authors amongst highest contributors in mutually cited journal when journal gets sudden peak in impact factor. Same author publishes in both journals and either mutually cites its own papers or his co-author\'s paper. Hence author self-citation and author co-author relations play a vital role towards such sudden inflation in impact factor. In a mutual citation case between \texttt{CoRR} and \texttt{TIT}, paper count of biasing journal abruptly increases in a period between 2007-2011. When we study author temporal pattern for this specific duration, we find an overlapping set of authors; Out of which, for biased journal whose impact factor largely increased H. Vincent Poor is found to be editor in chief for \texttt{TIT} for a period between 2003 to 2007, Syed A. Jafar is found to be associate editor for \texttt{TIT} from 2009 to 2011 and David Tse is a permanent member in fellowship for \texttt{TIT} journal \footnote{http://www.itsoc.org/people/committees/publications/2005}. While publishing more papers in \texttt{CoRR} journal and simultaneously, positioned as editorial board member in \texttt{TIT} journal authors either cite their own articles or papers of co-author. Such underlying author co-author and author editor relations influence large impact factor for biased journal such as \texttt{TIT} in this case. 

Publishers and editorial board members for commercial motive try to publicize and gain high impact factor for their journal. For all patterns, we find a close nexus between publishers which contribute towards influencing large citations. Six major publishing houses including {\em Elsevier, IEEE, ACM, Springer, Oxford Press, Wiley and others} involve in self-citation (see Figure \ref{fig:selfcitationph}) and mutual citation patterns of which {\em IEEE} and {\em Elsevier} make highest contributions. We find that 38.09\%, 35.71\% and 21.42\% cases are tagged with same publication house in high, medium and low weighted mutual citation buckets respectively. Extending it to chain and triangle relations, we find 
17.85\%  and 8.3\% relations with same publication house. Here also, {\em IEEE} and {\em Elsevier} are sole contributors. Combined with inherent domain specific nature; publishers come up with occasional review journals which are published annually or sister journals and give large citations to its parent journal which largely inflates impact factor. New journals belonging to same or different publication houses are also prone towards such patterns.

Citation is a spontaneous process where, degree of excursus is not directly determinable. Several factors play along in its manipulation. Here, we detect several bi-directional patterns and tried to understand the intention behind such sudden superfluous outflow in citations. It is found that mostly such patterns are field specific and highlights intrinsic property of a publication house to mutually uplift performance of all its journals. However, a rapid increase in paper count inflating impact factor of its associate journal is a possible indication towards anomalous citation activity. Such grouping in journal-journal citation network  is influenced by underlying editor co-authorship networks, author-author collaboration networks and a multilayer approach to take into account all such possible reasons. Abrupt increase in paper count, author self-citation, influence of scholarly relations (author editorial board relations, author co-author relations, influence of publication houses play vital role behind such relations.

\section*{Conclusion}
\label{S:7}
The aim of this paper is to discover various anomalies that exists between journals. Using citation data from Microsoft Academic Search, we extract bi-directional patterns and try to understand the intention behind such sudden superfluous outflow in citations. It is found that mostly such patterns are field specific and highlights intrinsic property of a publication house to mutually uplift performance of all its journals. A rapid increase in publication count inflating impact factor of its associate journal could be a possible indication towards anomaly. Author self-citation is also a significant reason that uplifts impact factor of journal on  a macroscopic scale. Moreover, such grouping in journal-journal citation graph is influenced by underlying author co-author, editor-co-author, author-author collaboration networks. An immediate future study would be to design an automated system that can take citation pattern early of a journal's career and predict if the journal experiences any anomalous citation pattern. One would further be interested to incorporate this anomalous citation pattern into the calculation of impact factor and refine it. 


\section*{Author Contributions}
\begin{itemize}
\item{\bf Conceptualization:} Tanmoy Chakraborty, Dinesh  Pradhan, Joyita Chakraborty, Subrata Nandi.
\item {\bf Data curation:} Tanmoy Chakraborty.
\item {\bf Formal analysis:} Dinesh  Pradhan, Joyita Chakraborty, Hridoy Sankar Dutta.
\item{\bf Investigation:} Tanmoy Chakraborty, Dinesh  Pradhan, Joyita Chakraborty, Subrata Nandi.
\item{\bf Supervision:} Tanmoy Chakraborty, Subrata Nandi.
\item {\bf Writing --  original draft:}  Dinesh  Pradhan, Joyita Chakraborty.
\item {\bf Writing – review \& editing:} Tanmoy Chakraborty.
\end{itemize}

\section*{Acknowledgments}
Tanmoy Chakraborty would like to thank the support of the Infosys center for Artificial Intelligence and Ramanujan Faculty Fellowship.


\end{document}